\documentclass[a4paper, 12pt]{article}

\def \be {\begin{equation}}
\def \ee {\end{equation}}
\def \bea {\begin{eqnarray}}  
\def \mea {\nonumber\\}
\def \eea {\end{eqnarray}} 
\def \UO {{\widehat U}} 
\def \AO {{\widehat A}}
\def \BO {{\widehat B}}
\def \HO {{\widehat H}}

\usepackage{amssymb}
\usepackage{mathabx}
\begin{document}
\title{A Geometric Substructure for  Quantum Dynamics} 
\author {A. J. Bracken
\\
Centre for Mathematical Physics 
\\School of 
Mathematics and Physics
\\University of Queensland\\
Brisbane, Australia\\
 }
\date{ }
\maketitle 

\begin{abstract} 
The description of a closed quantum system is extended with the 
identification of an underlying
substructure enabling an expanded formulation of dynamics in the Heisenberg picture.
Between measurements a ``state point" moves  
in an underlying  multi-dimensional complex projective space with constant 
velocity determined by the quantum state vector. Following a measurement 
the point changes 
direction and moves with new constant velocity along one of 
several possible new orthogonal paths with probabilities determined by the Born Interpretation of the state vector.
From this previously hidden substructure a new picture of 
quantum dynamics and quantum measurements emerges without violating existing no-go
theorems regarding hidden variables. A natural generalisation to a Riemannian 
substructure is proposed, determined by the entropy of the background environment.
This leads to a suggested
interaction between the substructure of quantum dynamics 
and the background gravitational field. 
\end{abstract} 
\vskip15mm
\noindent
{\bf email:} ajb@maths.uq.edu.au
\vskip3mm
\noindent

\vskip3mm
\noindent
{\bf ORCID:} 0000-0001-5181-4483 

\newpage

\section{Introduction}
There have been many attempts to describe aspects of quantum mechanics in 
geometric and sometimes also information-theoretic terms.  
See in particular (\cite{wiener} - \cite{nielsen2}) 
and references therein. Meanwhile efforts to identify ``hidden variables" that 
clarify the nature of the quantum measurement process have 
been restricted by powerful 
``no-go" 
theorems \cite{vonneumann, bell}.  The difficuties facing the quantisation of 
general relativity have also received a great deal of attention over many 
years -- see for example 
\cite{carlip} and references therein -- with no agreement
that any satisfactory resolution has been achieved, leading to the suggestion 
that rather than trying to quantise general relativity it may be more sensible to ``gravitise" quantum mechanics
\cite{penrose1,penrose2}. 

The
description of the measurement process and the associated 
Born interpretation of the state vector
have been the most contentious features of quantum mechanics since its inception.
In the case of a conservative system the strangeness 
of the orthodox description 
is seen most clearly in the Heisenberg picture
of
quantum dynamics \cite{vonneumann, dirac, bohm}.
There 
the state vector $|\psi\rangle$ is a constant unit vector in a Hilbert space $\cal{H}$, 
possibly infinite-dimensional, between measurements at times 
$t_0$ and $t_1>t_0$,  while self-adjoint operators 
$\AO(t)$, $\BO(t)$, $\dots$ representing observables  evolve in time
in accordance with Heisenberg's equation of motion
\bea
i\hbar \frac{d \AO}{dt}=  [\AO\,,\HO]\,,\quad{etc.}
\label{evolution1}
\eea 
Here $\HO$ is the Hamiltonian operator. (Only  
closed systems for which $\HO$ is time-independent are considered at this stage.)

But in another and  very different type of dynamical process that is 
assumed to  reflect
the interaction of the system with the measuring apparatus and the observer, 
a maximal set of commuting observables is chosen for  measurement  
and the state 
vector is subsequently  observed to 
move into a common eigenvector of that set
with a probability determined by 
Born's Rule \cite{vonneumann, dirac, bohm}. 
(Complications 
that arise with observables having
partially continuous spectra are set aside here.)

Suppose that at time $t_0$ a measurement has been made of  such a
set of commuting observables 
and 
$|\psi\rangle$ has been observed subsequently 
in one of their common eigenvectors.  The state vector
remains constant for $t>t_0$ until another measurement takes place.   
We may expand it during such a time interval  in terms of some 
chosen reference basis,
a complete set of orthonormal vectors
$|\varphi_i\rangle$, $i=1\,,2\,\dots\,$, possibly infinite in number, 
with constant complex expansion coefficients
$\alpha^i$.  Then
\bea
|\psi\rangle =\sum_i \alpha^i|\varphi_i\rangle
\quad {\rm with} \quad
\sum_i|\alpha^i|^2= \langle\psi|\psi\rangle=1\,.
\label{basis2}
\eea

Now consider another, in general different maximal set of commuting 
observables $\AO(t)$, $\BO(t)$, $\dots$ evolving in time 
for $t>t_0$ in accordance with (\ref{evolution1}) and having a 
complete orthormal set of common eigenvectors
$|\chi_ {K}(t)\rangle$ for  $K=1\,,2\,,\dots$
with components 
$\beta_{K}^{\,\,\, i}(t)$ in the same 
reference basis used for $|\psi\rangle$, so
\bea
|\chi_{K}(t)\rangle=\sum_i \beta_{K}^{\,\,\,i}(t)|\varphi_i\rangle\,.
\label{chiexpand}
\eea
(In the special case when $\HO$ is one of the chosen set then all the observables in the set are constants of the motion and their common eigenvectors $|\chi_ {K}\rangle$ and all their components $\beta_K^{\,\,\,i}$ are constant.)

Note that orthogonality of $|\chi_K(t)\rangle$ and $|\chi_L(t)\rangle$ for $K\neq L$ implies
\bea
\sum_i \overline{{\beta}_{K}^{\,\,\,i}(t) }\beta_{L}^{\,\,\,i}(t)=0\,,\quad K\neq L\,,
\label{orthogonality2}
\eea
where the overbar indicates  complex conjugation.

Suppose that a measurement is made at some time  $t_1>t_0$ of this second set of observables.  
Immediately prior to this second measurement the state vector can be expressed as

\bea
|\psi\rangle = \sum_K 
\langle \chi_K{(t_1)}|\psi\rangle |\chi_K{(t_1)}\rangle\,,
\label{expansion}
\eea
and the system can be said to be in one of the states $|\chi_K(t_1)\rangle$ 
in this superposition, which one being
indeterminate until the measurement is completed.  
Immediately  following the measurement, for each $K$ there is
according to Born's Rule  a probability 
\bea
P_K=|\langle \chi_K(t_1)\psi\rangle|^2=\sum_i 
|\overline{\beta_{K}^{\,\,\,i}(t_1)}\alpha^i|^2
\label{born2}
\eea 
of an observer finding the system with a new constant state vector equal to
$|\chi_K(t_1)\rangle$. 
If for example the system is observed to be in the state with vector 
$|\chi _{K'}(t_1)\rangle$ say, then for $t>t_1$ 
\bea
|\psi\rangle=|\chi_{K'}(t_1)\rangle=\sum_i \beta_{K'}^{\,\,\,i}(t_1)|\varphi_i\rangle
\label{born3}
\eea
and
\bea
\AO (t_1)\,
 |\psi\rangle  = \AO (t_1) |\chi_{K'}(t_1)\rangle=
a|\chi_{K'}(t_1)\rangle=a |\psi\rangle\,,
\mea
\BO (t_1)\,
 |\psi\rangle  =  \BO (t_1) |\chi_{K'}(t_1)\rangle=
b|\chi_{K'}(t_1)\rangle=b\,  |\psi\rangle\,, \quad\dots
\label{eval1}\
\eea
for some corresponding eigenvalues $a$, $b$, $\dots$  

If no observation is made of the state following the measurement, the 
system sits 
in the mixture of pure states $|\chi_K(t_1)\rangle$ for $t>t_1$ 
with associated probabilities $P_K$ as in (\ref{born2}).

To summarise this standard description of quantum dynamics in the 
Heisenberg picture: between the measurements at times $t_0$ and $t_1$ 
the operators in the 
set  to be measured at $t_1$ evolve in time in accordance 
with (\ref{evolution1}).  Their associated eigenvectors 
$|\chi_K(t)\rangle$ also evolve accordingly 
as do their expansion coefficients $\beta_K^{\,\,\,i}(t)$, which 
may be pictured as a group of quantities ``rotating" unitarily while 
remaining orthogonal as in (\ref{orthogonality2}) until the measurement at $t_1$. 
Then the system is observed to move into 
a new state with state vector 
$|\psi\rangle=|\chi _{K'}(t_1)\rangle$ say,  and the corresponding coefficients 
$\beta_{ K'}^{\,\,\,i}(t_1)$ are selected from the group. 
Both $|\psi\rangle$ and its expansion coefficients
$\beta_{ K'}^{\,\,\,i}(t_1)$ remain constant thereafter -- until another measurement, perhaps. 

\section{Identifying a substructure}

The reader may observe that the dynamical process described above  
is strongly reminiscent of the behaviour of a free 
particle travelling with constant velocity between impulsive forces
applied at $t_0$ and $t_1$. 
This suggests the association of the state vector $|\psi\rangle$ with
the velocity vector of a point moving in a hitherto unidentified underlying space. 
Accordingly the
$V\,\alpha^i$ are now identified with
the components  $v^i$ of the constant velocity vector  of a ``state point"
moving in an underlying space $S$ say, with complex coordinates 
$z^i$ so that
\bea
v^i=\frac{d z^i}{dt}\,,\qquad i=1\,,2\,,\,\dots
\label{veloc1}
\eea
(For convenience a constant $V$ with dimensions of 
velocity 
 $LT^{-1}$ has been  inserted here so that each $z^i$ has dimensions of length $L$.)
Note from (\ref{basis2}) that
\bea
\sum_i \overline{ v^i} v^i = V^2\,.
\label{velocity1}
\eea

Suppose that the state point starts at a location with coordinates 
$z^{\,\,i}_{0}$ at time $t_0$.  Because the $v^i$ are constants it follows trivially 
from (\ref{veloc1}) that
\bea
z^i(t)=z^{\,\,i}_{0}+v^i\,(t-t_0)\,,\quad t_0\leq t\leq t_1\,.
\label{zmotion}
\eea

For $t>t_1$ the  point moves in a new direction determined by 
which of the eigenvectors $|\chi_K(t_1)\rangle$ 
results from application of Born's Rule (\ref{born2}) to the measurement at $t_1$.  
The possible directions are determined by 
and associated with the corresponding 
``velocity vectors"   having components 
$w_K^{\,\,\,i}(t_1)=V\,\beta_{K}^{\,\,\,i}(t_1)$ with 
$\beta_{K}^{\,\,\,i}(t_1)$ as in (\ref{chiexpand}).  
Note that these directions are 
orthogonal according to (\ref{orthogonality2}) and that each velocity  vector is
normalized as in (\ref{velocity1}).
Note also that the probabilities associated with the different directions 
as given by  (\ref{born2})  can be viewed as
the moduli squared of generalized direction cosines between the velocity 
vector immediately before the measurement $v^i$ and
those possible immediately after the measurement, the $w_K^{\,\,\,i}(t_1)$, since
\bea
\sum_i |\overline{v^i(t_1)}w_K^{\,\,\,i}(t_1)|^2=
V^2 \sum_i  |\overline{\alpha^i}\beta_{K}^{\,\,\,i}(t_1)|^2=V^2 P_K\,.
\label{gen_cosines1}
\eea

For $t_0<t<t_1$ the vectors $w_{K}^{\,\,\,i}(t)$ evolve in time as noted above and  
may be pictured as a 
set of orthogonal velocity vectors rotating unitarily about the state point 
as it moves along the straight line (\ref{zmotion}).  
If the state vector observed following the measurement at $t=t_1$ is 
$|\chi_{{ K'}}(t_1)\rangle$ then for $t>t_1$ the state 
point moves with new constant velocity $w_{ K'}^{\,\,\,i}(t_1)$, so that
\bea
z^{i}(t)=z^{\,\,i}_{0}+v^{i}\,(t_1-t_0)+w_{ K'}^{\,\,\,i}(t_1)\,(t-t_1)\,,\quad t>t_1\,.
\label{zmotion2}
\eea
 
For $t<t_1$ it is indeterminate which of the evolving 
velocity vectors $w_{K}^{\,\,\,i}(t)$
will result from the measurement at  $t_1$. 
If no observation is made the new state for $t>t_1$ is a mixture of the eigenvectors 
weighted by the probabilities given by Born's Rule and the trajectory
of the state point belongs to a ``fan" of possible trajectories weighted accordingly.  

Note that when the choice of reference basis is altered by a unitary transformation
\bea
|\varphi'_i\rangle=\UO|\varphi_i\rangle\Rightarrow|\varphi'_i\rangle= \sum_{j} U_{i}^{\,j}|\varphi_j\rangle\,,\qquad\qquad
\mea\mea
 \qquad U_{i}^{\,j}(U^{\dagger})_{j}^{\,k}
=(U^{\dagger})_{i}^{\,j}U_{j}^{\,k}
=\delta_{i}^{\,k}\,,\quad (U^{\dagger})_{i}^{\,j}=\overline{U_{j}^{\,\,i}}\,,
\label{unitary1} 
\eea
any coordinate basis in $S$ with elements $z^i$ must 
undergo the corresponding unitary transformation
\bea
z'^{\,i}=\sum_{j} U_{j}^{\,i}\,z^{j}\,.
\label{unitary2}
\eea

In addition to $\UO$, the action of all linear operators on the Hilbert 
space of state vectors can be extended to action as matrices on the state point and its velocity in $S$, for example
\bea
\AO (t)|\psi\rangle \Rightarrow \sum_j A_{j}^{\,\,i}(t)\,v^j\,,\quad 
	A_{j}^{\,\,i}(t)=\langle \varphi_i|\AO (t)|\varphi_j\rangle
\label{operators2}
\eea
and if as before $\AO (t)$, $\BO (t)$ $\dots$ 
comprise the set of commuting operators  measured at $t=t_1$
and the system moves into the state $|\chi\rangle_{K'}(t_1)$ after the measurement, 
then
\bea
 \sum_j A_{j}^{\,\,i}(t_1)\,w_{ K'}^{\,\,\,j}(t_1)= a\, w_{ K'}^{\,\,\,i}(t_1)\,,
\mea
 \sum_j B_{j}^{\,\,i}(t_1)\,w_{ K'}^{\,\,\,j}(t_1)= b\, w_{ K'}^{\,\,\,i}(t_1)\,,\dots
\label{eval2}
\eea
 corresponding to (\ref{chiexpand}) and (\ref{eval1}).

Note also that because any state vector $|\psi\rangle$ can be 
identified with $e^{i\theta} |\psi\rangle$ for 
every real $\theta$, the space $S$ can be assumed to have 
the projective property that any two points with coordinates $z^{i}$ and 
$e^{i\theta}z^{i}$ for all $i=1\,,2\,,\dots $ are to be identified for every real $\theta$.

At this point it is worth emphasizing that the quantum system as described may consist
of arbitrarily many interacting particles (or subsystems). 
Accordingly, the Hilbert space $\cal{H}$ could be 
the tensor product of many Hilbert subspaces.  

More generally the tensor product of as many Hilbert subspaces as there 
are quantum systems can be considered 
whether interacting with each other or not.  The space associated 
with a particular system of interest
can accordingly be considered to be a subspace of a much larger Hilbert space. 
The underlying space $S$ is enlarged accordingly. 
What is essential  for the present discussion is that the encompassing Hilbert space
$\cal{H}$  has a countable basis as in (\ref{basis2}).  The $z^i$ then label points in 
the enlarged space $S$. 

Unlike the closed system on which attention has been focussed so far 
some of these extra systems will be open and 
interacting continuously or intermittently, unitarily or non-unitarily 
with their environment rather than at isolated points associated with measurements. 
They enter increasingly complicated mixed states 
as time passes \cite{sudarshan} - \cite{breuer}.

(These extensions of the description to include 
multiple quantum systems, whether interacting or not and whether open or not,
are not  pursued here.  Furthermore the complications arising from the introduction of 
quantum fields are not considered.) 

The generalised description of one closed quantum system as described above
can be thought of as an extension of the ``matrix mechanics" 
formulation of quantum dynamics \cite{green} which is as old as
quantum mechanics itself and has its origins in the pioneering work of 
Heisenberg and Born.  Observables are represented there by Hermitian matrices,
time-dependent in general, and defined as above. 
All quantum mechanical calculations can now be carried out in terms of the
state point and its velocity.  For example the expectation value of an 
observable $A(t)$ in the state $|\psi\rangle$ between measurements can
be expressed as
\bea
\langle A(t)\rangle = \sum_{i\,,j}\overline{\alpha^{j}}A_{j}^{\,\,i}(t_1)\alpha^{i}
=(1/V^2)\, \sum_{i\,,j}\overline{v^{j}}A_{j}^{\,\,i}(t)v^{i}
\label{expec_value}
\eea 
What is new is that the extension describes a previously hidden
substructure that provides a different way of thinking about
quantum dynamics and the quantum measurement process, 
as described in the next section.

\section{Hidden variables and  quantum measurement} 

Are the hitherto unrecognized variables $z^{i}$ the much discussed 
``hidden variables"
that resolve long-standing questions about
quantum indeterminacy and quantum measurement more generally?  The short answer
is ``No."  There has been extensive discussion of these questions since 
the birth of quantum theory -- 
see for example \cite{vonneumann, bohm}, \cite{schrodinger} - \cite{kochen} 
and especially the decisive work \cite{bell}.
  
As described above,
the state point has a fan  of possible future directions to 
choose from at $t=t_1$ which is
converted to a mixture of uncertain outcomes by the
measurement process in accordance with  Born's Rule.  
The role of the observer \cite{green2} is to convert this resultant
mixed state into a pure state by identifying which of the possible 
trajectories the state point follows
after the measurement.  

Before the measurement the Shannon - von Neumann 
entropy of the system has the value
\bea
-\rho \log(\rho) =0
\label{entropy1}
\eea
where $\rho$ denotes the density matrix \cite{vonneumann},  
which has just the one
eigenvalue $1$ when the system is in a pure state $|\psi\rangle$ and
\bea
\rho=|\psi\rangle\langle \psi|\,.
\label{entropy2}
\eea
Then the state point 
has a definite trajectory as in (\ref{zmotion}).  
After the measurement at $t=t_1$
\bea
\rho=\sum_K P_K |\chi_K(t_1)\rangle\langle\chi_K(t_1)|
\label{entropy3}
\eea 
and the entropy takes the larger value
\bea
\sum_K P_K\log(1/P_K)\,.
\label{entropy2}
\eea
Here the $P_K$ are the Born probabilities of the trajectories 
in the fan as in (\ref{born2}).
The entropy  decreases to $0$ again after the observation of which of those trajectories the
state point now follows, with the system again in a pure state.  

Only simple measurements are considered here as described in
\cite{vonneumann,dirac,bohm} and by many others since.  More sophisticated
descriptions of the measurement process and more generally of a quantum system's
possible interactions with its environment
have been developed over the years  -- see in particular
\cite{sudarshan,kraus,breuer,wiseman} -- together with more sophisticated
measures of the system's entropy \cite{ellinas}. 

The reader may consider that the behaviour of the state point before and after a measurement
is analogous to 
that of a macroscopic object floating down a horizontal stream that forks into 
two such streams at right angles.  The square of the direction cosine between the
direction of either fork and that of the original stream may be considered a first estimate 
of the probability that the object will float down that particular fork.
For example if the two forks are at angles of $\pi/6$ and $\pi/3$ with the 
original stream, with associated direction cosines $\sqrt{3}/2$ and $1/2$  
respectively, then the associated probabilities are $3/4$ and $1/4$.

The critical difference between this classical behaviour and that associated with the
quantum measurement is that the indeterminacy in the behaviour of the 
classical object prior to reaching the fork can in principle be reduced arbitrarily greatly by
more refined observation of the system
so that it becomes more certain which fork
will be followed.   Except in special cases \cite{bell} 
this is not possible in the quantum case.  
In short, the classical indeterminacy prior to the fork is arbitrarily reducible in principle 
whereas in general the quantum indeterminacy prior to the measurement -- associated
with the set of velocity vectors rotating unitarily about the trajectory of the state point --  is irreducible.

The quantum measurement itself is now to be 
regarded as an interaction of the
quantum system with its macroscopic environment at the point in $S$ 
reached at time $t_1$.  
The state point moves between such points associated with 
measurements at times $t_0$ and $t_1$. 

\section{A suggested generalization}

Several questions suggest themselves.
What is the nature of the points in $S$ associated with measurements?
More generally what interpretation can be given  to the  space $S$
in which the process underlying quantum dynamics occurs?  
Why does the state point move in a straight line between measurements? 
 
It is convenient to address these questions in the context of a natural generalizarion of
the dynamical substructure described so far and 
it is now proposed that the space underlying quantum dynamics as described above
is actually a locally flat subspace of a more general space $S$, a
complex Riemannian manifold with associated Hermitian metric tensor
\bea
g_{i\,j}(\bar{z}\,,z)=
\overline{g_{j\,i}({\bar{z},\,{z})}}\,.
\label{metric1}
\eea
Here $z$ denotes the point in $S$ with coordinates $z^i$, and $\bar{z}$ 
its complex conjugate. 
Infinitesimal distance-squared on the manifold is then  defined as
\bea
ds^2=g_{i\,j}\,d\bar{z}^id{ z}^j=\widebar{ds^2}\,,  
\label{riemann1}
\eea
where the summation convention has now been introduced.

In general the points in $S$ associated with measurements
of a closed quantum system can be described as  local singularities or ``stagnation points" in $S$
associated with the location in space-time where the measurements take place,
being typically 
the location of  measuring devices in a meta-stable state \cite{green} -- 
think of a cloud chamber, a Geiger counter or a photographic plate, for example.

As to the meaning of $S$, it is proposed that it represents the entropy 
content (equivalently, the information content) 
of the physical environment within which quantum systems 
evolve, 
including any measuring devices. 
This implies that the structure of $S$ changes when 
the entropy content of the environment changes.
For example when a photographic plate is exposed during a quantum measurement it is
clear that the entropy of the neighbouring environment increases abruptly  
as the associated singularity in $S$ disappears.
More generally ``measurements" may simply refer to 
interactions between quantum systems and 
their environment whether continuous or stochastic, at singular points 
in $S$ or non-locally.  In the absence of  observations  quantum systems
move after such interactions into  more and more complicated
mixed states with higher entropy.  It is suggested that while such interactions 
increase the entropy of the quantum systems involved and their environment, they also 
provide the (only) source for changes in the structure of $S$. 
 
One possible description of that structure  would be provided by 
supposing that it is a K\"ahler manifold \cite{molitor,nakahara}
with real potential $B(\bar{z},z)$ such that
\bea
g_{i\,j}(\bar{z}\,,z)=\frac{\partial ^2 B(\bar{z},z)}{\partial {\bar z}^i \partial{z}^j}\,,
\label{kahler1}
\eea
and by supposing further that  $B$ is a measure of the entropy/information content
of the environment in which quantum systems evolve.  It would remain to determine a  mathematical description of
how changes in $B$ arise from  interactions of those systems with their environment. Note that (\ref{riemann1}) and  (\ref{kahler1}) would imply
\bea
ds^2=\frac{\partial ^2 B(\bar{z},z)}{\partial {\bar z}^i \partial{z}^j}\,d{\bar z}^i\,dz^j\qquad
\mea \rm{and}\qquad\qquad\qquad\qquad\qquad\qquad\qquad\qquad
\mea
\frac{\partial g_{i\,j}}{\partial z^k}=\frac{\partial g_{i\,k}}{\partial z^j}\,,\quad \frac{\partial g_{i\,j}}{\partial \bar{z}^k}=\frac{\partial g_{i\,k}}{\partial \bar{z}^j} \,.
\label{kahler2}
\eea

Bearing (\ref{kahler1}) in mind as a possibility, while returning to the general 
discussion of the behaviour of 
the state point  of a closed system 
between measurements, it is clear that
straight line motion as in (\ref{zmotion}) is naturally  generalized to motion
along a geodesic between 
the locations at $P$ and $Q$ say,  of quantum measurements, 
so minimizing the distance travelled.
This may be regarded as an analogue of the principle of least action that 
leads to geodesic motion of a mass point (a ``test particle") 
in space-time  \cite{bergmann} and leads here to  
a variational condition in the familiar form \cite{synge}
\bea
0=\delta\,\int_{P}^{Q}\,ds=\int_{u_P}^{u_Q} \left(g_{i\,j}
\bar{p}^{i} p^{j}\right)^{1/2}\,du\,,\quad \bar{p}^{i}=\frac{d{\bar z}^i}{du}\,,\quad
 p^{j}=\frac{dz^j}{du}\,,
\label{variation1}
\eea
where $u$ is a parameter measuring distance along the geodesic.  

It then follows by a generalization from the real \cite{synge} 
to the complex case that
\bea
g_{i\,j}\,\frac{dp^j}{du}
-\frac{\partial g_{l\,k}}{\partial{\bar  z}^i}\,\bar{p}^l p^k
+\frac{\partial g_{i\,j}}{\partial{\bar  z}^k}\bar{p}^kp^j
+\frac{\partial g_{i\,j}}{\partial z^k} p^jp^k
+\quad{\rm{c.\,c.}}=0\,.
\label{variation2}
\eea

Supposing that the metric on $S$ is non-singular 
with inverse $g^{i\,j}$ such that
\bea
g^{i\,j}g_{j\,k}=\delta^i_{\,k}=g_{k\,j}g^{j\,i}\,,
\label{inverse1}
\eea
an absolute derivative  of $p^m$ with respect to $u$ along a geodesic can be
defined from (\ref{variation2})  by 
\bea
\frac{\delta p^m}{\delta u}=g^{m\,i}\left(g_{i\,j}\,\frac{dp^j}{du}
-\frac{\partial g_{l\,k}}{\partial {\bar z}^i}\,\bar{p}^l p^k
+\frac{\partial g_{i\,j}}{\partial {\bar z}^k}\bar{p}^kp^j
+\frac{\partial g_{i\,j}}{\partial z^k} p^jp^k\right)
\mea\mea
=\frac{dp^m}{du}-g^{m\,i}\left(\frac{\partial g_{l\,k}}{\partial {\bar z}^i}\,\bar{p}^l p^k
-\frac{\partial g_{i\,j}}{\partial{\bar z}^k}\bar{p}^kp^j
-\frac{\partial g_{i\,j}}{\partial z^k} p^jp^k\right)
\label{absolute1}
\eea
together with its complex conjugate.  The vanishing of $\delta p^m/\delta u$
along a geodesic then leads to $z^m(u)$ by integration given (\ref{variation1}).

The vector (with components)  $v^i$ corresponding to the the state vector $|\psi\rangle$  is now to be considered as parallel transported
along the geodesic
traced by $z^i(t)$, where $u$ is now replaced by elapsed time $t$ 
along that geodesic and $p^i(u)$ ($=dz^i/du$)  
is replaced by $v^i(t)$ ($=dz^i/dt$).  The state vector and 
corresponding velocity vector $v^i$  are no longer 
constant between measurements.  Instead   
 $v^i$   has  
vanishing absolute derivative along the geodesic as defined from
(\ref{absolute1}), so that 
\bea
0= \frac{\delta v^m}{\delta t}=
\frac{dv^m}{dt}-g^{m\,i}\,\left(\frac{\partial g_{l\,k}}{\partial {\bar z}^i}\,\bar{v}^l v^k
-\frac{\partial g_{i\,j}}{\partial {\bar z}^k}\bar{v}^kv^j 
-\frac{\partial g_{i\,j}}{\partial z^k} v^jv^k\right)\,.
\label{absolute2}
\eea

Note that all of (\ref{metric1})-(\ref{variation1}) and 
(\ref{inverse1})-(\ref{absolute2}) are consistent with
the projective condition introduced in Sec. 2 which 
identifies $z^i$ with $e^{i\theta} z^i$ for all values of $i$ and every 
real $\theta$ and which now
carries over to the generalisation of $S$ to a Riemannian manifold, 
whether K\"ahler or not. 

The Hilbert space $\cal{H}$  is now taken 
to be the tangent space to the geodesic followed by 
$z^{i}(t)$, obtained by integrating (\ref{absolute1}). 
Vectors in $\cal{H}$, including the quantum state vector $|\psi\rangle$, 
are parallel transported along the geodesic, 
preserving lengths and orthogonality relations. 

It is natural to assume further that the Hamiltonian operator 
$\HO$ has vanishing absolute derivative 
along the geodesic, obtained by regarding its matrix representation
$H_{j}^{\,\,i}$ as a mixed tensor and generalizing (\ref{absolute2})  
accordingly \cite{synge}, 
while 
the governing 
differential equation for other operators representing
time-dependent observables includes an extra term  
generalizing (\ref{evolution1}).

As before, following a measurement the state vector moves into a new vector 
among the eigenvectors of the
set of commuting operators being measured in accordance with Born's Rule
and accordingly $z^i(t)$ embarks along a new geodesic.

The differences between the simple substructure described in 
the preceding sections and 
the generalized substructure of the quantum dynamics proposed here can be expected to
have implications 
for the outcome of quantum measurements and for quantum 
dynamics more generally, not least 
because of possible effects of the curvature of $S$, in particular on the 
evolution of the state vector in accordance with
(\ref{absolute2}).  
More analysis  is necessary to determine 
the nature of such implications.

Consideration of the 
space $S$  suggests a further generalization with possibly greater
consequences
for physics, as we discuss in the next section.

\section{Interaction with the gravitational field}
As mentioned in the Introduction there has been extensive discussion
over many years of attempts to quantise the theory of general relativity, and
more recently of the possibility of ``gravitising" quantum theory as an alternative
approach to resolving the disconnect between the two theories, each of which boasts 
major successes in its own domain.  
The identification of a Riemannian space underlying quantum dynamics suggests a
different resolution of this problem, one which treats quantum theory and
the theory of relativity on a more equal footing, and leads to  the following final proposals:

$\bullet$ The geometric space $S$ underlying quantum systems 
can be considered jointly with space-time carrying the 
local gravitational field, with combined  coordinates 
$(z^i, x^{\mu})$, for $i=1\,,2\,,\dots$ and 
$\mu = 0\,,1\,,2\,,3\,,$ and combined  metric tensor
and infinitesimal distance-squared
\bea
g_{i\,j\,\mu\,\nu}(\bar{z},z,x)\,,\quad d\sigma^2=
g_{i\,j\,\mu\,\nu}d\bar{z}^idz^j dx^{\mu}dx^{\nu}\,.
\label{newmetric1}
\eea
Here $x^{\mu}$ are the usual space-time coordinates. 

$\bullet$ Interactions of quantum systems with their environment are
labelled not only by the coordinates $z^i$ 
corresponding points in $S$ but also by the space-time 
coordinates $x^{\mu}$ of the point or points in space-time at which such interactions
occur. 

$\bullet$ The metric tensor is not in general a simple product
\bea
g_{i\,j\,\mu\,\nu}\neq g_{i\,j}g_{\mu\,\nu}\,,
\label{newmetric2}
\eea
in particular during measurements and perhaps also during more general
interactions between quantum systems and their environment.  This implies  
that the local gravitational field interacts with 
measurement processes in particular and may influence quantum dynamics  more generally -- see for example 
 \cite{penrose2} for related discussions in other contexts.   It implies also that 
the changing entropy content of the
space $S$ during such processes  can alter the local gravitational field.  
In short, changing entropy at the quantum level 
can be an unexpected 
source of gravitational field strength -- a kind of ``dark energy."  How this new type
of interaction affects Einstein's equation for the gravitational field and equations governing change to the structure 
of $S$ must be the subject of further study.  

\section{Concluding remarks}

The simple substructure  identified in Sec. 2 provides a new way of thinking 
about  quantum dynamics and measurements without suggesting any
new observable effects.  On the other hand, the generalizations suggested 
in the following sections would surely have far-reaching and important implications 
for physics. Further study is encouraged.
\vskip3mm
\noindent
{\bf Acknowledgement:} The author is grateful for the mentorship provided over
many years by H.S. Green, A.O. Barut and L. Bass., and thanks W.M.J. Bracken, D. Ellinas and P.D. Jarvis for helpful comments.

\end{document}